\documentclass[aps,prl,reprint,groupedaddress,nofootnoteinbib]{revtex4-2}

\usepackage{graphicx}
\usepackage{braket}

\begin{document}
\newcommand{\tr}{\mathop{\rm Tr}\nolimits}
\newcommand{\figsize}{\small}


\title{Tuning to the Edge of the Abyss in SU(5)}


\author{Howard Georgi}
\email[]{hgeorgi@fas.harvard.edu}
\affiliation{Center for the Fundamental Laws of Nature\\
Jefferson Physical Laboratory\\
Harvard University, Cambridge, MA 02138
}


\date{\today}

\begin{abstract}
I show that 
if a dimensionless parameter is tuned to be close to the boundary of the positivity
domain and symmetry breaking is driven by a cubic term in the Lagrangian,
the scale of the physics of 
symmetry breaking in a quantum field theory as measured
by the Higgs mass
can be much greater than the dimensional scales in the classical Lagrangian.
Radiative corrections produce large and physically important corrections,
helping to stabilize the large VEV.  The resulting picture contrasts
sharply with the ``modern'' view of QFT as an effective field theory.
I describe how this
mechanism might produce the GUT scale in an SU(5) model in which the
dimensional parameters in the Lagrangian are at the low scale.
\end{abstract}


\maketitle


Dimensional analysis is one of our simplest and most powerful tools.  
But there are situations in which it can be confusing if
used uncritically.  
In this note, I show that in a field theory
in which a dimensionless 
parameter can be tuned close to the boundary of the positivity
domain, a cubic term in the Lagrangian can produce symmetry breaking
in which the scale of the physics of symmetry breaking is
parametrically larger than the dimensional scales in the classical
Lagrangian.  This is not exactly a failure of dimensional analysis because
there is a small dimensionless parameter measuring the distance to the
boundary.  However, it 
is an outlandish suggestion because it does not fit with the
modern picture of quantum field theory as an effective theory.
We have gotten used to thinking about quantum field theory as a low-energy
approximation to some other high-energy physics. 
It is hard to argue with this if we are looking at a field theory of
quasiparticles in condensed matter physics.  But for fundamental physics,
until we understand quantum gravity or string theory or whatever comes
next, I believe that this jury is still out and we should consider
unconventional possibilities like the one proposed here.   

For a renormalizable
Lagrangian, the classical positivity domain is the region of the space of
dimensionless parameters
for which the quartic potential of scalar fields
$\phi$ is positive definite so that the potential is positive for large
$|\phi|$.  On the boundary of the positivity domain, some homogeneous 
combination of
components of $\phi$ vanishes as a coupling $\lambda\to0$. 
If we are very near that boundary, the
quartic potential depends very weakly on that combination. Then cubic 
terms in the potential
may push the minimum to very negative
values for large $|\phi|$ so that some components of the VEV are
parametrically large 
compared to the dimensional parameters, with the Higgs mass and other
physical parameters going to infinity as we go to the
boundary.  Cubic terms are crucial.

The application I will describe is to a toy GUT model.  
But 
before I describe the model, it may be helpful to explain why the standard
model is \textbf{not} an example of the mechanism.  In an ungauged
doublet Higgs model with quartic coupling $\lambda$ and negative mass
squared $-m^2$, the VEV squared is $v^2\propto |m^2|/\lambda$ which goes to
$\infty$ as $\lambda\to0$. However
the Higgs mass squared, which is the scale of the physics of symmetry breaking,
is $\propto\lambda v^2\propto |m^2|$ which remains
fixed and of the order of the dimensional parameter in the Lagrangian as
$\lambda$ gets small.  Of course, if we gauge the electroweak 
symmetry to get the standard model, the gauge boson masses get large as
$\lambda\to0$, but at least at tree level, their masses are just
proportional to the gauge couplings which have nothing to do with the
physics that produces the VEV.\footnote{The Coleman-Weinberg mechanism,
\cite{Coleman:1973jx}, is a special case in which the gauge couplings are
crucial to the physics of symmetry breaking.  This mechanism will play an
important role here in radiative corrections, but our symmetry breaking is
driven by the cubic term in the potential which is absent in Coleman-Weinberg.}

In the mechanism I will describe, the symmetry breaking is driven not by a
negative mass-squared term ($m^2$ can be positive or zero), but by a cubic
term in the Lagrangian with 
coefficient $\kappa$ with dimensions of mass rather than mass squared.  
As a coupling $\lambda$ goes to zero, the VEV of a
field goes like $v\propto\kappa/\lambda$ 
and the tree-level Higgs mass goes to $\infty$
 as $\lambda\to0$ like $\kappa/\sqrt{\lambda}$ for fixed $\kappa$.
And there
are even heavier particles with mass 
$\propto|\kappa|/\lambda$, which are important in the
radiative corrections.\footnote{We will see below that radiative corrections make the
physical Higgs mass larger, scaling like the heavier particle masses.}
Thus there are large physical scales in the
theory that are independent (at tree level) of the gauge couplings without
any large scales in the Lagrangian.  

When I decided to investigate 
this mechanism,
I rather expected that quantum corrections would drastically change
things. And indeed, I
find that the leading quantum
corrections are large.  But they actually stabilize the vacuum with the
large VEV.

I believe that this
mechanism can be applied to many models with trilinear couplings and
multiple quartic
couplings.  A relatively simple example where a
mechanism to produce 
a large VEV might be physically interesting is an SU(5) gauge
theory~\cite{Georgi:1974sy} with a 24 of 
scalars\footnote{This model, sometimes including the cubic term in the
potential, has been studied for over 
50 years.  It is possible that somewhere in the extensive literature on
the subject, there is a discussion of the crucial effect of the cubic term
when a quartic coupling goes to the positivity boundary.  If so, I
apologize for missing it and  
I assume that the authors will enlighten me.}.  We consider a toy GUT
with only a 24 of scalars.  It is not crazy to think of this by itself
because the 24 cannot couple directly to the standard model fermions, so it
makes some sense to ignore the fermions in the physics of the GUT symmetry
breaking.   In the
conventional picture, the large VEV 
of the 24 is generated by a conventional Higgs mechanism with a negative
mass-squared term of the order of the square of the GUT scale.  But in fact, we
can tune close to the abyss and let the 
vacuum do this job with no large dimensional parameters in the Lagrangian.
Take the 24 scalar field, $\phi$, to be a hermitian, 
traceless 5$\times$5 matrix, and write the most general invariant
potential%
\begin{equation}
\begin{array}{c}
V(\phi)=\frac{3\lambda_{14}}{50}\Bigl(30\tr(\phi^4)-7\tr(\phi^2)^2\Bigr)
\\+\frac{3\lambda_{23}}{100}\Bigl(13\tr(\phi^2)^2-20\tr(\phi^4)\Bigr)
\\+\sqrt{\frac{10}{3}}\,\kappa\tr(\phi^3)+\frac{m^2}{2}\tr(\phi^2)
\end{array}
\end{equation}%
This has the required cubic interaction and $\lambda_{14}$ and
$\lambda_{23}$ define the domain of positivity.
The coefficients of $\lambda_{14}$ and $\lambda_{23}$ are both positive
semi-definite.  The $\lambda_{14}$ term vanishes when
\begin{equation}
\begin{array}{c}
\phi\propto U\,\beta_{23}\,U^\dagger
\mbox{~~where~~}
\\ {[\beta_{23}]}_{jk}\equiv \frac{1}{\sqrt{30}}
\left(2\delta_{jk}-5(\delta_{j4}\delta_{k4}
+\delta_{j5}\delta_{k5})\right)
\end{array}
\label{23}
\end{equation}
for some unitary $U$.
The $\lambda_{23}$ term vanishes when
\begin{equation}
\begin{array}{c}
\phi\propto U\,\beta_{14}\,U^\dagger
\mbox{~~where~~}
\\ {[\beta_{14}]}_{jk}\equiv \frac{1}{\sqrt{20}}
\left(\delta_{jk}-5
\delta_{j5}\delta_{k5}\right)
\end{array}
\label{14}
\end{equation}
Thus positivity requires $\lambda_{14}>0$ and $\lambda_{23}>0$ and if one
of these two couplings becomes very small, the $\kappa$ term drives the
corresponding component of  $\phi$ to a very large VEV.  For very small
coupling, the effect of the mass term is very small, and a non-zero mass
complicates the formulas enormously, so for pedagogical reasons we will
take $m=0$ in the rest of the paper.  This is not a renormalizable
constraint because $m^2$ depends on the renormalization scale which we
will discuss later.  But it simplifies the formulas without
making any essential difference for small $\lambda$.

In this limit, the VEV is either (\ref{23}) or
(\ref{14}). The results are summarized below.
\begin{equation}
\begin{array}{c}
V(\phi_{23}\beta_{23})
=\frac{1}{12}\,\phi_{23}^3(3\lambda_{23}\phi_{23}-4\,\kappa)
\\
\braket{\phi_{23}}=\frac{\kappa}{\lambda_{23}}\quad\quad
V(\braket{\phi_{23}}\beta_{23})
=-\frac{\kappa^4}{12\lambda_{23}^3}\rule{0ex}{2ex}
\end{array}
\label{vev23}
\end{equation}
\begin{equation}
\begin{array}{c}
V(\phi_{14}\beta_{14})
=\frac{1}{4}\phi_{14}^3(3\lambda_{14}\phi_{14}-2\sqrt{6}\,\kappa)
\\
\braket{\phi_{14}}=\sqrt{\frac{3}{2}}\,\frac{\kappa}{\lambda_{14}}
\quad\quad
V(\braket{\phi_{14}}\beta_{14})
=-\frac{9\kappa^4}{16\lambda_{14}^3}\rule{0ex}{1.5ex}
\end{array}
\label{vev14}
\end{equation}

In the application to GUTs we would like to have
$\lambda_{23}<2^{2/3}\lambda_{14}/3$
so that the $\beta_{23}$ vacuum is picked out and the symmetry breaks to
SU(3)$\times$SU(2)$\times$U(1). 
We will discuss this case in detail.  
The potential $V(\phi_{23}\beta_{23})$ 
has an inflection point at $\phi_{23}=0$ and for $\kappa>0$ and small
$\lambda_{23}$, there is a deep minimum at $\kappa/\lambda_{23}$ as shown in
figure~\ref{fig-1}.  
{\figsize\begin{figure}[htb]
$$\includegraphics[width=1\hsize]{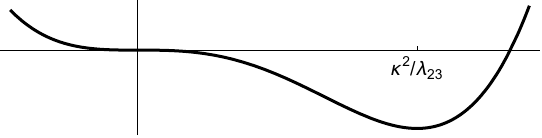}$$
\caption{\figsize\sf\label{fig-1}The shape of the tree-level
potential}\end{figure}}For 
non-zero real $m$, the inflection point splits (barely noticeably for small
$\lambda_{23}$) into a shallow local minimum at
$\phi_{23}=0$ and a weak local maximum for $\phi_{23}$ of order $m^2/\kappa$
which stays fixed and irrelevant when $\lambda_{23}\to0$ as the deep
minimum moves out to infinity. 

Then the massive particles in the theory are the
$(3,2)$ of leptoquark gauge bosons, a scalar $(1,1)$ which I will refer to
as the GUT-Higgs to avoid confusion with the standard model Higgs, and the
adjoint scalars transforming like $(8,1)$
and $(1,3)$ under SU(3)$\times$SU(2).  Their tree-level masses for small
$\lambda_{23}$ are
{\renewcommand{\arraystretch}{1.5}
\begin{equation}
\begin{array}{|c|c|c|c|}
\hline
(3,2)&(1,1)&(8,1)&(1,3)\\
\hline
\displaystyle \frac{g\kappa}{\lambda_{23}}&
\rule[-3ex]{0ex}{8ex}
\displaystyle \frac{\kappa}{\sqrt{\lambda_{23}}}&
\displaystyle \sqrt{\frac{6\lambda_{14}}{5}}\,\frac{\kappa}{\lambda_{23}}&
\displaystyle \sqrt{\frac{6\lambda_{14}}{5}}\,\frac{2\kappa}{\lambda_{23}}
\rule[-3ex]{0ex}{8ex}\\
\hline
\end{array}
\label{all}
\end{equation}}%
where $g$ is the gauge coupling.
 All have masses that go to infinity
as $\lambda_{23}\to0$ for fixed $\kappa$, but at different rates!  
The smaller mass for the GUT-Higgs would be very interesting if it persisted
beyond tree level because it would be an intermediate mass scale, but I
will argue that the one loop corrections modify this 
quantitative detail while leaving
intact the important qualitative feature of large physical masses in the small
$\lambda_{23}$ limit.

We can address the quantitative issue using the background
field method \textit{a la} Coleman-Weinberg~(CW)~\cite{Coleman:1973jx}
and constructing the CW potential as a function of $\phi_{23}$.
The SU(3)$\times$SU(2) 
ensures that the other fields in (\ref{all}) will be eigenstates 
of the $\phi_{23}$ dependent
``mass'' that appears in the CW calculation.  

The CW contribution to the potential is 
\begin{equation}
\frac{1}{64\pi^2}\sum_{j=1}^4n_jM_j(\phi_{23})^4
\log\frac{M_j(\phi_{23})^2}{\mu^2}
\label{cw}
\end{equation}%
where $\mu$ is an arbitrary renormalization scale 
and $n_j$ is the degeneracy of
the multiplet. The $M_j$ and $n_j$ are
shown in Table~\ref{mjs}.
{\begin{table}[htb]
$$\begin{array}{|c|c|c|c|}
\hline
j&\mbox{type}&n_j&M_j(\phi_{23})^2\\
\hline
1&\mbox{(3,2)}&36*&g^2\,\phi_{23}^2\\
2&\mbox{(1,1)}&1&\phi_{23}(3\lambda_{23}\phi_{23}-2\kappa)\\
3&\mbox{(8,1)}&8&\phi_{23}
\bigl((6\lambda_{14}+3\lambda_{23})\phi_{23}/5+4\kappa\bigr)\\
4&\mbox{(1,3)}&3&\phi_{23}
\bigl((24\lambda_{14}-3\lambda_{23})\phi_{23}/5-6\kappa\bigr)	\\
\hline
\multicolumn{4}{c}{\mbox{*There are 12 color-flavor states,
each with 3 spins.}}
\end{array}
$$
\caption{\sf\label{mjs}The $\phi_{23}$-dependent masses that appear in
(\ref{cw})}\end{table}} 

Taking a cue from \cite{Coleman:1973jx}, we can simplify the analysis by
choosing $\mu$ so that the CW term leaves the VEV at its classical
value.  This
requires
{\renewcommand{\arraystretch}{3}\begin{equation}
\log \mu^2=
\frac{\sum_jn_jM_j^2(\phi_{23}){M_j^2}'(\phi_{23})\left(\log
M_j^2(\phi_{23})+1/2\right)}
{\sum_jn_jM_j^2(\phi_{23}){M_j^2}'(\phi_{23})}
\end{equation}}%
In the small $\lambda_{23}$ limit, 
this gives a $\mu$ of the order of the mass of the heaviest particles
(scalar or gauge boson).

The resulting CW contribution to the effective potential can be much larger than the
tree level contribution but it further stabilizes the tree-level vacuum.
For some typical parameters, an example is shown in 
figure~\ref{fig-2}.
{\figsize\begin{figure}[htb]
$$\includegraphics[width=1\hsize]{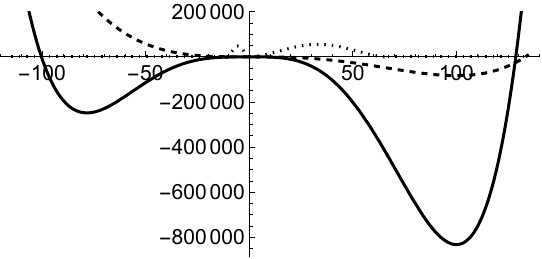}$$
\caption{\figsize\sf\label{fig-2}$\Re(V_{\rm eff}(\phi_{23})$ including the CW
contribution (solid line) for $\lambda_{14}=0.3$,
$\lambda_{23}=0.001$, $\kappa=1$, and $m=0$.  
The dashed line is
tree-level.  The dotted line is $10^4\times\Im(V_{\rm eff})$.}\end{figure}}
The general shape of the corrections is easy to understand from (\ref{cw})
and Table~\ref{mjs}.  All of the $M_j(\phi_{23})^2$ are proportional to
$\phi_{23}$ (this is exact in our approximation of $m^2=0$, but in
general the $M_j(\phi_{23})^2$ 
are all small at $\phi_{23}=0$).  And they are 
$\propto\phi_{23}^2$ for large $\phi_{23}$.  This means that the CW
contribution has at least an approximate quadratic zero at 
the origin.  This is a local maximum because the logs are negative.  By
construction the CW contribution has an extremum at
$\phi_{23}=\kappa/\lambda_{23}$. And because of the two $\phi_{23}$s in
the mass squares, it grows like $\phi_{23}^4$ at large
$\phi_{23}$.  The minimum scales like $\kappa^4/\lambda_{23}^4$, so for
sufficiently small $\lambda_{23}$, it dominates the tree-level contribution.
One can think of figure~\ref{fig-2} approximately as a symmetric CW potential 
\begin{equation}
\propto \phi^4\Bigl(\log(\phi^2/\mu^2)-1/2\Bigr)
\end{equation}
tilted by 
the effect of the cubic term.

This also means that the CW
contribution to the GUT-Higgs mass squared scales like
$\kappa^2/\lambda_{23}^2$ and for 
sufficiently small $\lambda_{23}$ will overwhelm the tree level contribution
that scales like $\kappa^2/\lambda_{23}$, so the scale of the physics of symmetry 
breaking is proportional to
$1/\lambda_{23}$ like the scalar and massive gauge boson masses.  The CW
contribution to the GUT-Higgs mass squared for small 
$\lambda_{23}$ is
\begin{equation}
\frac{9(25g^4+56\lambda_{14}^2)\kappa^2}{50\pi^2\lambda_{23}^2}
\label{cwhiggs}
\end{equation}

Because the CW contribution to the GUT-Higgs mass squared is so large,
one might worry that higher order
contributions will continue to grow and the loop expansion will be
useless.  I do not expect this to happen.  
The point is that dimensional analysis implies that large physical masses are
 proportional to the large VEV which is proportional to $1/\lambda_{23}$.
The tree level GUT-Higgs mass squared is proportional to
$1/\lambda_{23}^2$ from the VEV squared times a coupling factor of
$\lambda_{23}$. 
The point is that the 1-loop corrections are
not anomalously large.  The tree-level result is anomalously small because
of the small coupling factor $\lambda_{23}$ (which is not present for the
other heavy particles in (\ref{all}).
The only way
the small $\lambda_{23}$ coupling can get into a denominator is through the
VEV, (\ref{vev23}), so dimensional analysis guarantees that squared masses
beyond tree level
will be proportional to $1/\lambda_{23}^2$ time a power series in
the larger coupling, $\lambda_{14}$, as usual in perturbation theory.

It is a long way from this toy SU(5) model to a realistic GUT.
Because the symmetry-breaking mechanism is different, some
model-building issues will have to be solved in new and different ways.
But I hope that I have convinced the reader that this mechanism for
symmetry breaking with cubic terms and small couplings at the edge of the
positivity domain is worth exploring as a possible 
alternative to the conventional scheme.
I don't believe that this solves
the tuning problems in GUTs, but it might replace the usual tuning problems
with very different ones.  
One might even dream of a theory in which the large scale of gravity arises in
this way.  And I somehow just like the idea that some of the
puzzles in fundamental physics might arise because our world is 
perched precariously near the edge of oblivion.

\section*{Acknowledgements}
I am grateful to David B. Kaplan, Ken Lane, Jackie Lodman, Lisa Randall, 
Matthew Schwartz and Cumrun Vafa for
helpful comments. 
This project has received support from the European Union's
Horizon 2020 research and innovation programme under the Marie 
Skodowska-Curie grant agreement No 860881-HIDDeN.

\bibliography{cw}
\end{document}